\documentclass[journal,10pt]{IEEEtran}
\usepackage{amssymb}
\usepackage{amsmath}
\usepackage{cite}
\usepackage{url}
\usepackage{xcolor}
\usepackage{subfigure}
\usepackage{cite,graphicx,amsmath,amssymb}
\usepackage{subfigure}
\usepackage{fancyhdr}
\usepackage{float}
\usepackage{mdwmath}
\usepackage{mdwtab}
\usepackage{caption}
\usepackage{amsthm}
\usepackage{setspace}
\usepackage{algorithm}
\usepackage{algorithmic}
\usepackage{makecell}
\usepackage{diagbox}
\usepackage{stfloats}
\usepackage{float}
\usepackage{picinpar}
\usepackage[colorlinks,
linkcolor=black,
anchorcolor=black,
citecolor=black
]{hyperref}
\allowdisplaybreaks
\makeatletter
\def\ScaleIfNeeded{%
\ifdim\Gin@nat@width>\linewidth \linewidth \else \Gin@nat@width
\fi } \makeatother
\usepackage{booktabs}
\usepackage{multirow}
\usepackage{graphicx}%图片

\usepackage{epsfig}
\usepackage{stfloats}%图片位置跨栏
\usepackage[T1]{fontenc}
\allowdisplaybreaks[4]

\begin{document}
\title{Near-field Integrated Sensing and Communication: Opportunities and Challenges}

\author{Jiayi Cong, Changsheng You, \IEEEmembership{Member, IEEE}, Jiapeng Li, Li Chen, \IEEEmembership{Senior Member, IEEE}, 
	
	 	Beixiong Zheng, \IEEEmembership{Senior Member, IEEE}, Yuanwei Liu, \IEEEmembership{Fellow, IEEE},  Wen Wu, \IEEEmembership{Senior Member, IEEE}, 
	
	Yi Gong, \IEEEmembership{Senior Member, IEEE}, 	Shi Jin, \IEEEmembership{Fellow, IEEE}, Rui Zhang, \IEEEmembership{Fellow, IEEE}  
\vspace{-2.9pt}	
	\thanks{J. Cong is with Southern University of Science and Technology and also with the Frontier Research Center, Peng Cheng Laboratory; C. You, J. Li and Y. Gong are with Southern University of Science and Technology; L. Chen is with University of Science and Technology of China;  B. Zheng is with South China University of Technology; Y. Liu is with Queen Mary University of London;  W. Wu is with the Frontier Research Center, Peng Cheng Laboratory; S. Jin is with the National Mobile Communications Research Laboratory, Southeast University; R. Zhang is with The Chinese University of Hong Kong, Shenzhen and National University of Singapore. \emph{(Corresponding author: Changsheng You)}}}

\maketitle

\begin{abstract}
With the extremely large-scale array (XL-array) deployed in future wireless systems,  wireless communication and sensing are expected to operate in the radiative \emph{near-field} region, which needs to be characterized by the \emph{spherical}  rather than planar wavefronts. Unlike most existing works that  considered far-field integrated sensing and communication (ISAC), we study in this article the new \emph{near-field ISAC}, which integrates both functions of sensing and communication in the near-field region. To this end, we first discuss the appealing advantages of near-field communication and sensing over their far-field counterparts, respectively. Then, we introduce three approaches for near-field ISAC, including joint near-field communication and sensing, sensing-assisted near-field communication, and communication-assisted near-field sensing. We discuss their individual research opportunities, new design issues, as well as propose promising solutions. Finally, several important directions in near-field ISAC are also highlighted to motivate future work.
\end{abstract}

\section{Introduction}

While the fifth-generation (5G) wireless systems are being globally deployed, researchers from both academia and industry have been exploring key technologies for future six-generation (6G) wireless systems, with the target of achieving more stringent and diversified performance requirements, such as super-high data rates, hyper-reliability and low-latency, ubiquitous connectivity.
Among others, \emph{extremely large-scale array/surface} (XL-array/surface) is a promising technology to greatly improve the spectral efficiency and spatial resolution, via  drastically  increasing the number of antennas in massive  multiple-input multiple-output (MIMO) by another order-of-magnitude (e.g., more than $256$) \cite{DaiMagazineNearfield1, WCMadd1}. Moreover, to harness  enormous spectrum resources, future wireless systems are migrating to higher frequency bands, such as millimeter wave (mmWave) and even Terahertz (THz) frequencies.

\begin{figure*}[t]
	\centering	
	\includegraphics[height=3.5in]{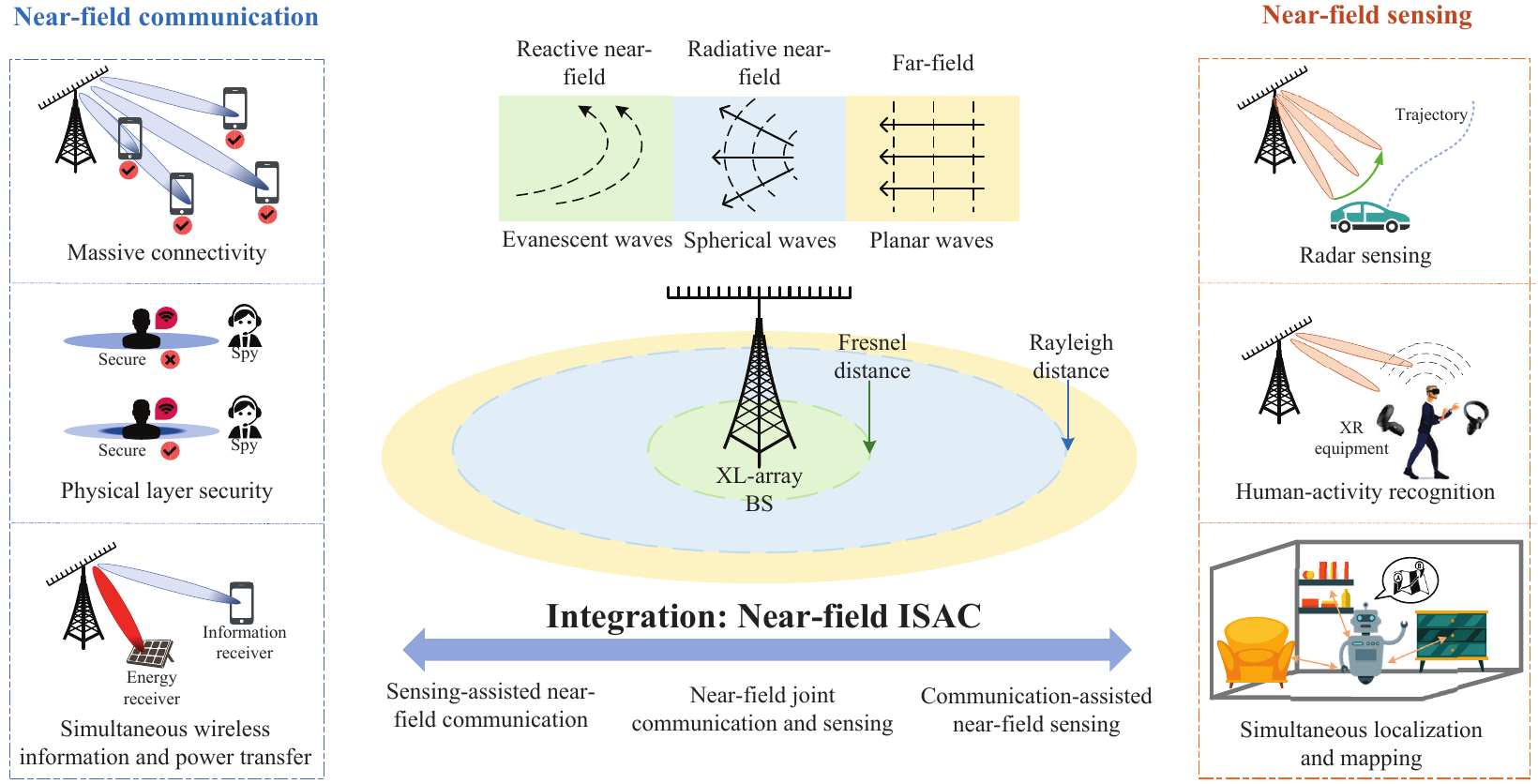}
	\caption{Typical application scenarios for near-field communication and sensing systems.}
	\label{fig_system}
\end{figure*}

The above technology trends lead to a fundamental change in the electromagnetic (EM) propagation environment, which shifts from the conventional far-field channel modeling to its \emph{near-field} counterpart. Specifically, for an XL-array receiver, its EM field can be roughly divided into three regions as shown in Fig.~\ref{fig_system}: 1) reactive near-field region, where the transmitter is within its \emph{Fresnel distance} and there are substantial amplitude and non-linear phase variations across the XL-array antennas; 2) radiative near-field region, where the transceiver distance is smaller than the so-called \emph{Rayleigh distance} that increases with the array aperture, for which the radio propagation is modeled by \emph{spherical} wavefronts; and 3) far-field region, where the transmitter is located beyond the Rayleigh distance and the phase variations can be approximated as linear ones \cite{LiuyuanweiMagazinenearfield2, YouMagazineNearfield3}. 

In particular, for the radiative near-field region (hereafter simply referred to as near-field), wireless channels are characterized by 
the spherical (instead of planar in the far-field region) wavefronts with spatial non-stationarity. Specifically,  the phase changes over the XL-array aperture are \emph{non-linear}, while the amplitude variations can be negligible or non-negligible depending on the transceiver distance. This new channel characteristic  potentially improves both  communication and sensing (C\&S) performance in the near-field as follows, with typical applications illustrated in Fig.~\ref{fig_system}.

\begin{itemize}
\item \textbf{Near-field communication:} First, in contrast to far-field beamforming that steers  beam energy along a certain direction,  near-field beamforming based on  spherical wavefronts  achieves a new function of \emph{beam-focusing}, which concentrates  beam energy in a given \emph{location/region} \cite{ZhanghaiyangBeamfocusingNearfield4}. This not only enhances the received signal power at the desired user{\footnote{In practice, the achieved array gain in the near-field is affected by the loss of the polarization mismatch across array antennas\cite{LiuyuanweiMagazinenearfield2}.}}, but also eliminates interference to undesired users. Consequently, users at different locations can be simultaneously served by the XL-array with small interference, hence allowing for massive connectivity.
Second, for the near-field line-of-sight (LoS) MIMO channel, its channel rank may be larger than one, thus potentially enhancing the spatial multiplexing gain \cite{wang2024tutorial}.
\item \textbf{Near-field sensing:} First, for near-field radar sensing, the spherical wavefront at the XL-array aperture can be utilized to estimate both the target angle and range by using e.g., the two-dimensional (2D) multiple signal classification (MUSIC) algorithm, thereby reducing the need for distributed arrays and their synchronization \cite{JinshiChannelNearfield5}. Besides, the enlarged array aperture provides finer-grained spatial resolution in both the angular and range domains. In addition, the near-field beam-focusing effect can be exploited to enhance the sensing signal-to-noise ratio (SNR) of echo signals for achieving more accurate estimation. While for other near-field sensing applications such as human-activity recognition, the spherical wavefront provides additional features in the range domain, hence potentially improving the recognition accuracy.
\end{itemize}

As wireless communication and sensing can be implemented in one system, it is natural to combine these two functions at the same infrastructure (e.g., base station (BS)), giving rise to the active field of \emph{integrated sensing and communication} (ISAC) \cite{LiufanISACand6G1}.{\footnote{ Compared to the C\&S in separate networks, the ISAC in federation networks enjoy the integration and coordination gains at the cost of higher design complexity.}} Compared with near-field communication and near-field sensing alone, near-field ISAC  provides new opportunities and challenges \cite{LYWISAC}. On one hand,  near-field beam-focusing effect can be exploited to mitigate inter-functionality interference. Besides, apart from the angular resolution in far-field ISAC, near-field ISAC endows another degree-of-freedom (DoF) in the range domain to achieve higher communication capacity and finer sensing resolution.  These advantages enable promising near-field ISAC applications, such as the high-resolution and location-aware extended reality (XR), and low-altitude economy with drones. However, on the other hand, it is also more challenging to develop efficient near-field C\&S integration methods due to the \emph{high-dimensional} spherical-wavefront channel modelling and new coupling between near-field C\&S  performance in different scenarios. This thus motivates the current work to investigate efficient approaches for near-field ISAC in high-frequency bands such as mmWave, including near-field joint C\&S (JC\&S), sensing-assisted near-field communication, and communication-assisted near-field sensing. We discuss their individual research opportunities, new design issues, and propose promising solutions to address them. Last, several interesting directions are  presented to motivate future work.

\section{Near-field Joint Communication and Sensing}
For near-field ISAC, one design paradigm is to jointly optimize both C\&S performance in a shared system architecture and hardware platform. Therefore, several 
new near-field effects have to be considered in balancing  C\&S performance tradeoff for both narrow-band and wide-band systems.
\subsection{Near-field JC\&S in Narrow-band Systems}

\begin{figure*}[!t]
	\centering
	\subfigure[Multi-beam design for near-field JC\&S.]{\includegraphics[height=2.0in]{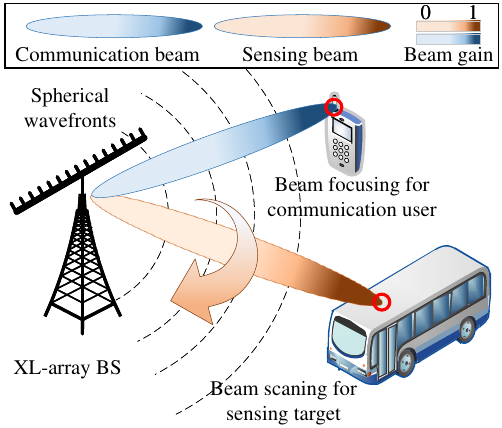}\label{fig_joint_narrow1}}%
		\hfil
	\subfigure[Near-field energy-spread effect.]{\includegraphics[height=2.2in]{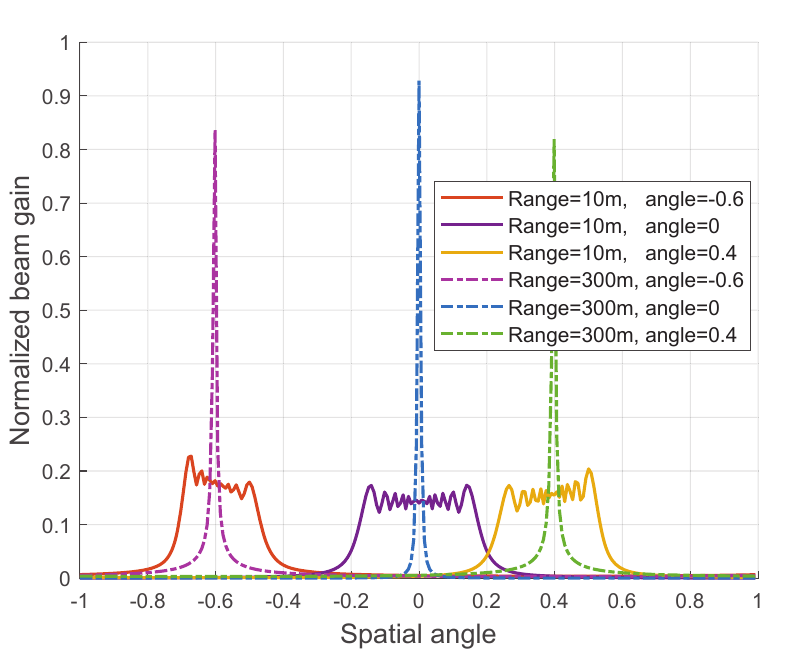}%
		\label{fig_joint_narrow2}}	
	\caption{Near-field JS\&C in narrow-band systems.}
	\label{fig_joint_narrow}
\end{figure*}

Consider a narrow-band near-field wireless system with its bandwidth much smaller than the carrier frequency. The BS is equipped with an XL-array to simultaneously serve multiple communication users and sense surrounding targets in its near-field region over the same bandwidth.  For JC\&S, efficient beamforming designs need to be devised to compensate for severe path-loss in high-frequency bands. Compared with time-division (TD) based beamforming that shifts communication and sensing beams in different time, it is more promising to concurrently generate multiple beams to improve both C\&S performance, as illustrated in Fig.~\ref{fig_joint_narrow1}. Specifically, to enhance the near-field communication performance at the target users (e.g., rate maximization),  communication beams usually are tuned  towards the communication users at \emph{fixed} locations/regions by exploiting the near-field beam-focusing effect. On the other hand, the beam control for near-field radar sensing generally depends on its detection/estimation aim. First, for target parameter (angle/range/Doppler) estimation, the sensing beams should dynamically scan the region of interest in both the angular and range domains. To balance the C\&S performance, 
one promising method is generating {multiple beams} in the near-field  for concurrent beam sweeping. However, it is worth noting that the conventional array-division based multi-beam design may not be directly applied to the near-field scenario, since different sub-arrays may observe different user angles and there exist coverage holes in the angular domain. To tackle these issues, an alternative and more efficient approach is by sparsely activating a portion of antennas of the XL-array to form an effective \emph{sparse linear array}, thus generating multiple beams simultaneously by exploiting both the main-lobe and grating-lobes, albeit at certain loss in the array gain.

Next, for target tracking, adaptive sensing beams need to be designed to track moving targets. In this case, both C\&S beams need to be directed towards the communication users and/or targets to improve C\&S SNR. The corresponding near-field C\&S performance is  determined by the \emph {subspace correlation} between C\&S channels \cite{LiufanISACandRis2magazine}. 
%Different from the far-field case, the correlation between near-field C\&S is more complicated. 
Generally speaking, there is weak channel correlation when  C\&S users are located at different angles and/or different ranges. This indicates that for near-field JC\&S, even if the targets and communication users are located at the \emph{same angle} but different ranges, multiple beams still need to be generated to achieve high C\&S SNR. This is in sharp contrast to the far-field case, where only one single beam is needed when they reside at the same angle \cite{LiufanISACandRis2magazine}. 
Furthermore, consider the more general \emph{mixed-field} scenario where  sensing targets are located in the near-field region of the XL-array, while communication users are located in the far-field. In this case, \emph{near-and-far} field channel correlations introduce new C\&S tradeoff \cite{YouMagazineNearfield3}. Specifically, due to the energy-spread effect shown in Fig.~\ref{fig_joint_narrow2}, when the near-field target angle is close to that of the far-field communication user, the far-field beam towards the communication user may yield strong signal power to the near-field target for enhancing its detection \cite{JinshiChannelNearfield5}; while the beam directed to near-field targets may also cause strong interference to the far-field communication users.
As a result, efficient power allocation and beam scheduling design can be developed to exploit/mitigate the near-and-far energy-spread effect.

\subsection{Near-field JC\&S in Wide-band Systems}
{For wide-band systems where the ratio between the bandwidth and carrier frequency is relatively large}, the near-field JC\&S design is more complex due to the more pronounced \emph{beam-split} effect. {An efficient approach is steering focusing-beams towards the communication users and sensing targets over different frequency bands, thus avoiding undesired interference even they are at the same angle or very close.} However, as the near-field beam-focusing location/region is generally affected by the operating frequency, beams at different frequencies in wide-band systems  are focused at different locations, thus affecting both  C\&S performance. 

%\begin{figure*}[htbp]
	%\centering	
	%\includegraphics[height=2.2in]{Fig_wide_band.pdf}
	%\caption{Beam-split control and hybrid beamforming architecture for near-field JC\&S.}
	%\label{Beam-split}
%\end{figure*}

For communication users, the near-field beam-split effect usually results in degraded received SNR, since the beam energy is distributed in different regions. To address this issue, a viable approach is to integrate true-time-delay (TTD) devices into XL-arrays to achieve \emph{frequency-dependent} beams \cite{elbir2023nba}. This is because TTDs can induce controllable time-delays to the signals, which become frequency-dependent phase shifts in the frequency domain. By contrast, for radar sensing, the near-field beam-split effect can be effectively exploited to enhance sensing performance \cite{DaiBeamaplitNearfield8}. Specifically, an efficient radar sensing method is utilizing the beam-split effect to generate multiple beams simultaneously, each focused at the same range but different angles. While in different time, the XL-array dynamically adjusts the focused ranges of multiple beams.
%, as illustrated in Fig.~\ref{Beam-split}. 
As such, the target angle and range can be easily resolved at the XL-array with low training/sensing overhead. For near-field JC\&S in wide-band systems, several design issues need to be addressed. For example, efficient algorithms need to be developed to improve the radar sensing performance (e.g., estimation Cram\'{e}r-Rao bound (CRB) minimization) under communication performance constraints (e.g., signal-to-interference-plus-noise ratio (SINR) maximization) in near-field scenarios. Note that there exists a fundamental tradeoff between TTDs allocation for C\&S given a fixed number of TTDs. Generally speaking, allocating more TTDs for radar sensing can expand the near-field sensing (spatial) coverage, while allocating more TTDs for  communications can more effectively aggregate beam energy at different frequencies to achieve enhanced received SNR. Therefore, one can jointly design efficient allocation of TTDs and optimize the TTD delays to balance the above tradeoff.

\section{Sensing Assisted Near-field  Communication}

In this section, we discuss efficient approaches for utilizing sensing information and/or wireless sensing methods to assist in near-field communication.

\begin{figure*}[t]
	\centering	
	\includegraphics[height=1.8in]{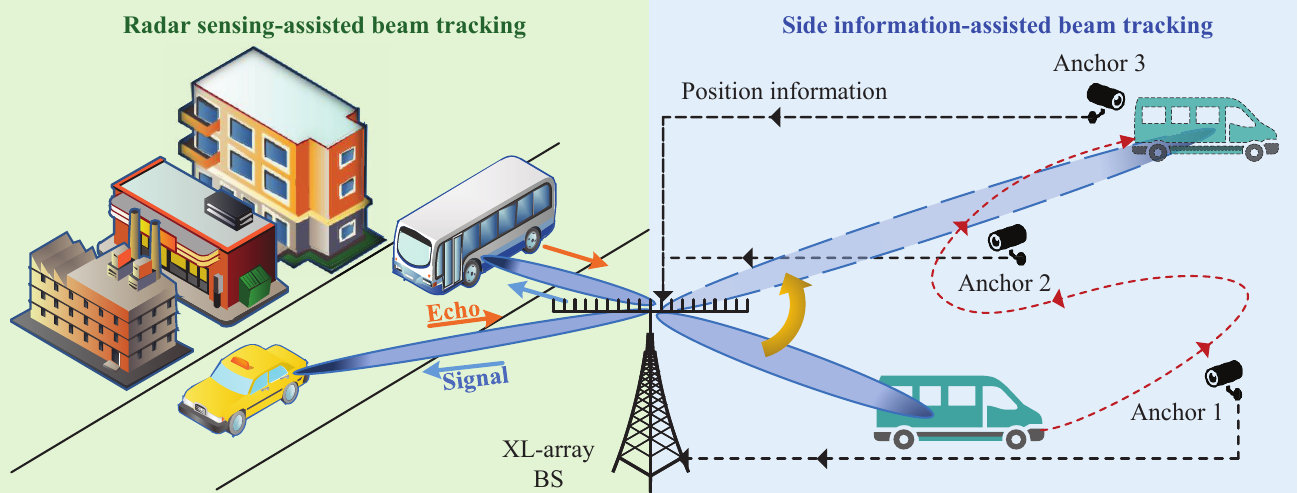}
	\caption{Sensing-assisted near-field beam training/tracking.}
	\label{Road}
\end{figure*}

\subsection{Sensing-Assisted Near-field Beam Training/Tracking}

For near-field communication, beam training and tracking are efficient approaches in practice to acquire necessary channel state information (CSI) for establishing initial high-SNR links. However, they generally require much higher training overhead than that in the far-field region, due to the required 2D beam search over both the angular and range domains.  Existing works on near-field beam training/tracking mainly reduce the overhead by designing various beam search methods (e.g., angle-then-range search, hierarchical search) \cite{WCMadd2}. Nevertheless, \emph{wireless sensing} has not yet been well exploited, which has the potential to substantially reduce the training overhead. Among others, one promising  approach is to exploit environmental sensing information to reduce the near-field beam search space. Specifically, one can deploy a number of environment anchors or sensors in the user zone or along the user trajectory, whose position information is known \emph{a priori} at the XL-array BS  and thus can be used as reference location information, as shown in Fig.~\ref{Road}. {If the anchors are deployed by different providers, a central node can be deployed (e.g., by the third party) to take in charge of the information fusion and sharing.} As such, at the beginning of near-field beam training/tracking, the anchors/sensors around the user notify the BS that the user is present within their sensing range.  Then the XL-array BS obtains an initial user location information based on the  positions of reporting sensors/anchors, and uses it to predict best beam angle/range.  Note that for this method, the required beam training/tracking overhead generally depends on the accuracy of initial user location information obtained. When it is sufficiently accurate, one can quickly determine the best user beam by simply searching the neighborhood of the initial location. However, if it deviates too far from the true user position due to e.g., too few sensors around the user, a large amount of training pilots is still needed when applying the exhaustive search in the neighborhood. Thus,  adaptive near-field beam training/tracking methods are required, so that  the optimal user beam can be accurately found with low training overhead, even when  the initial user location information is inaccurate.

Next, radar sensing methods can also be utilized to facilitate near-field beam tracking by exploiting echo signals reflected by  targets, as shown in Fig.~\ref{Road}. For this method, no dedicated downlink pilot signals are needed for beam tracking and  it also avoids feedback loops in communication-based beam tracking. 
Compared with the far-field case, the near-field radar-assisted beam tracking is  more challenging, since it tends to cause \emph{faster phase variations} change over the XL-array. Furthermore, robust radar sensing-assisted beam tracking needs to be designed for near-field scenarios, since near-field beams have very narrow beamwidths and are susceptible to  beam misalignment. An effective solution approach is dynamically controlling the near-field beamwidths according to the predicted user range and velocity for minimizing the beam misalignment probability, while maximizing the beam gain. In addition, one can also generate near-field multi-beams to cover the potential user locations along the predicted user trajectory.

\begin{figure*}[!t]

	\centering
	\subfigure[Global and local coordinate systems in coordinate transformation.]{\includegraphics[height=1.6in]{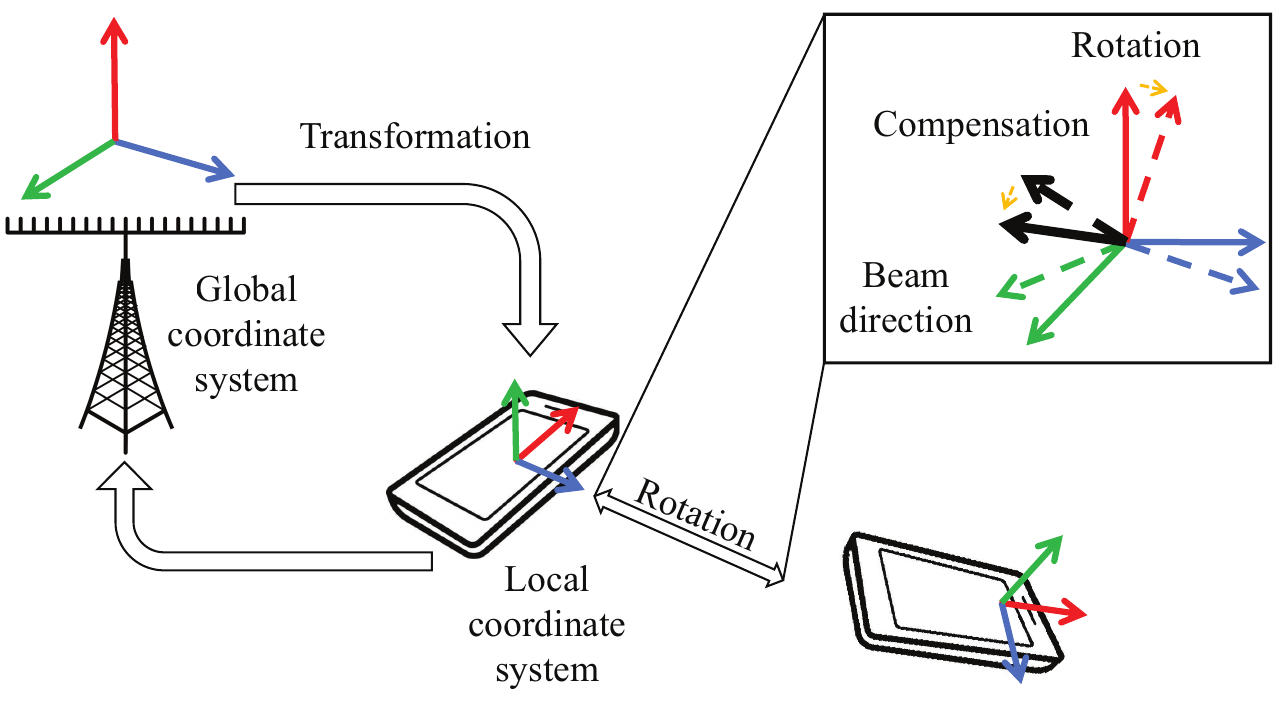}%
		\label{fig3b}} 
		\hfil
	\subfigure[Normalized array gain versus accuracy deviation of the sensor.]{\includegraphics[height=2.2in]{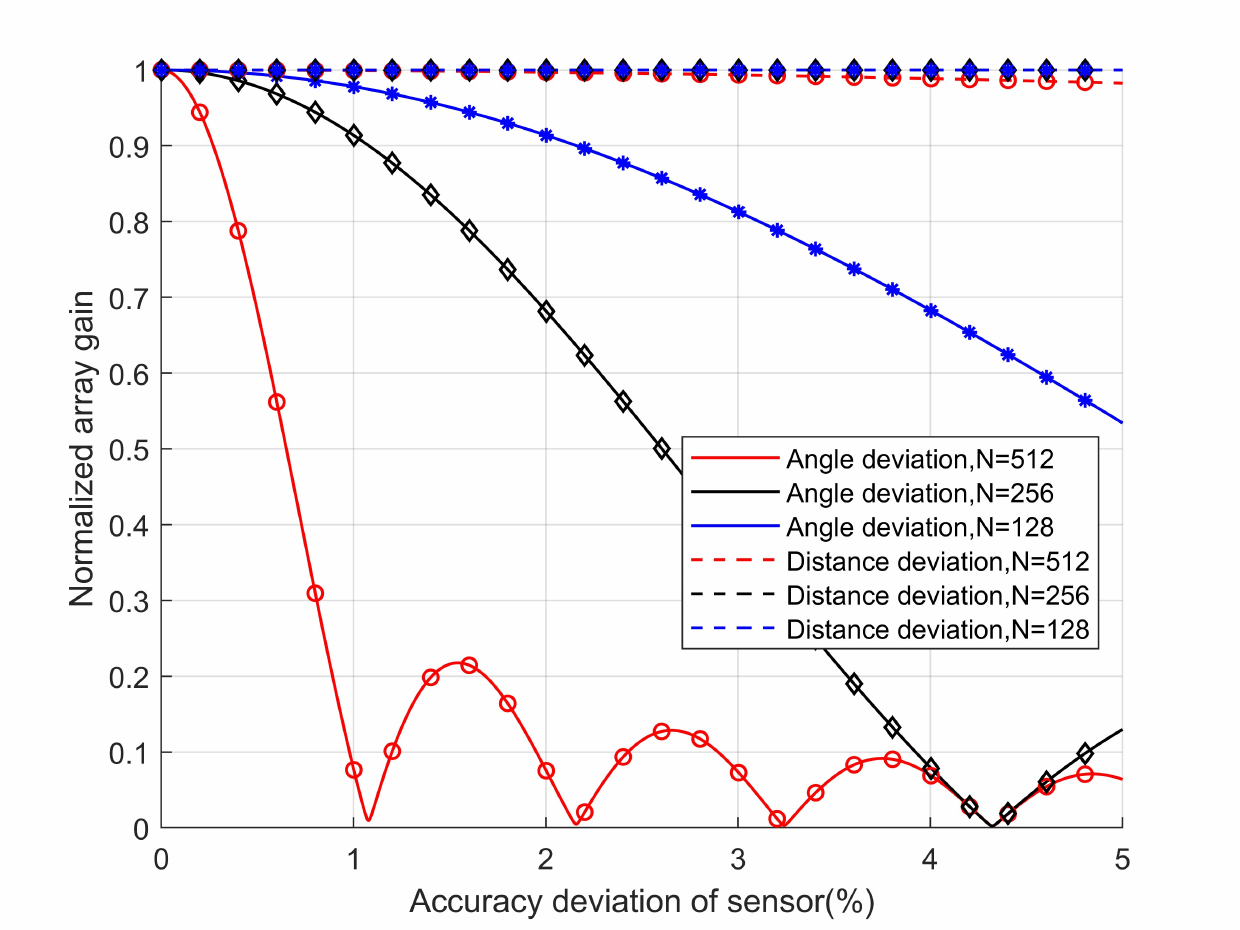}%
		\label{fig:4}} 
	\caption{Attitude information assist near-field beam focusing.}
	\label{fig:simo}
\end{figure*}
\subsection{Attitude Information-Assisted Near-field Beam Focusing}

In near-field scenarios, the user equipment (UE) may not be viewed as a point target, but rather  an extended target. In this case, in addition to position information, the \textit{attitude} information, which depends on the  physical shape of the UE, is also necessary for judicious beam control. For electronic devices, such as dipoles and wire antennas, the attitude parameter can be defined as the \emph{orientation}, which characterizes the spatial rotation angle. Note that even when the user is at fixed location, its UE rotation can still significantly impact the beamforming performance, as a slight  rotation can cause severe beam misalignment.
% in the polarization mismatch. 

The above issue can be addressed by exploiting both the position and attitude sensing information  in the near-field for precise beam-focusing. 
This can be effectively achieved by employing more accurate near-field channel modelling based on fundamental electromagnetic (Maxwell's) equations, such as electromagnetic propagation modeling (EPM), which essentially characterizes the functional dependence of received signals on the \textit{position} and \textit{attitude} parameters \cite{YouMagazineNearfield3}. Specifically, based on EPM, the signals observed on the receiving surface, such as electric fields, depend on the current distribution inside the dipole, which in turn is  determined by the position and attitude parameters.

Additionally, in mobile devices,  micro-electro-mechanical systems (MEMSs) also provide a cost-effective kinematic solution, where sensors such as tachymeters and gyroscopes can provide position and attitude information. 
The main procedures of near-field position and attitude information-assisted communication are as follows.
During each channel coherence time, the UE feeds back its position (e.g., angle, range) and attitude (e.g., rotation, displacement) information obtained by sensors to the BS. Such information is then used to control the beam direction and power for reducing the misalignment probability with low overhead, as shown in Fig.~\ref{fig3b}. To  this end, a coordinate transformation that takes into account position and attitude information needs to be established. For example, a global coordinate system originating at the BS can be used for controlling the transmit beamforming direction, while a local coordinate system, which is with respect to the mobile device that dynamically rotates, can be used for controlling the receive beamforming at the user.
Generally speaking, sensor accuracy plays a critical role in near-field  attitude information-assisted communication. As shown in Fig.~\ref{fig:4}, the array gain tends to decrease as the sensing error increases. Moreover, the greater the number of antennas, the more evident the near-field effect at the UE,  which results in greater array gain fluctuations. Furthermore, the gain degradation by angle deviation is more pronounced than that caused by range deviation, which indicates that  the UE rotation has a more significant impact than its displacement.

\begin{figure*}[!t]
	\centering
	\subfigure[Networked sensing with information fusion.]{\includegraphics[height=2.4in]{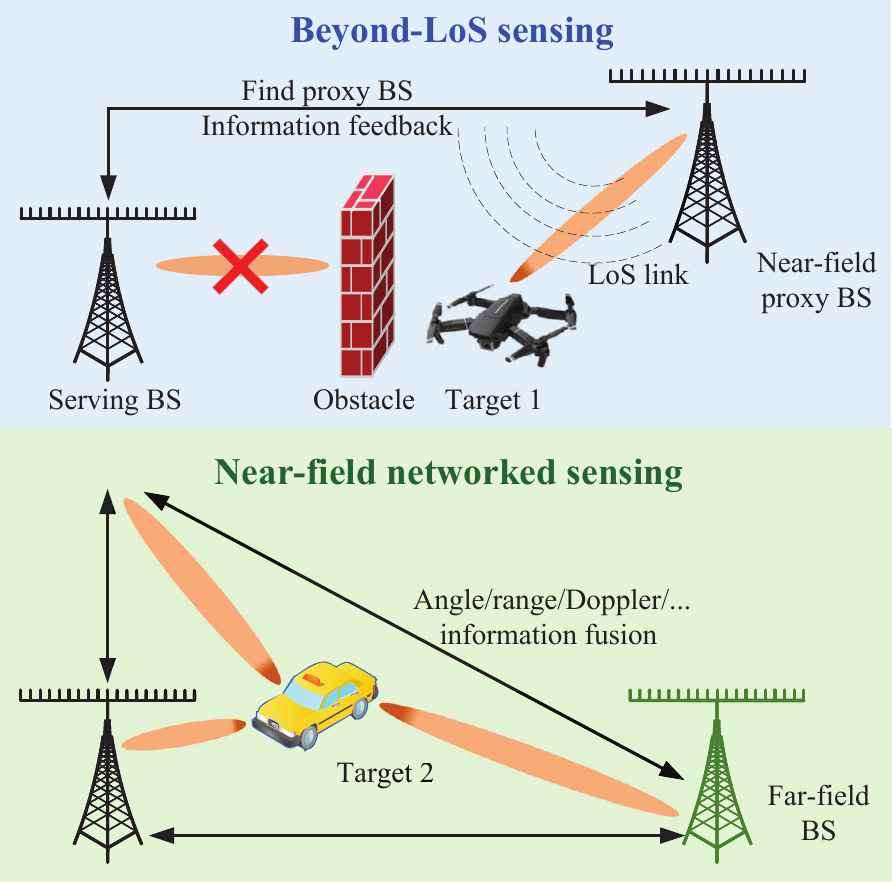}%
		\label{fusion_1}}
	\hfil
	\subfigure[CKM assisted near-field sensing.]{\includegraphics[height=2.4in]{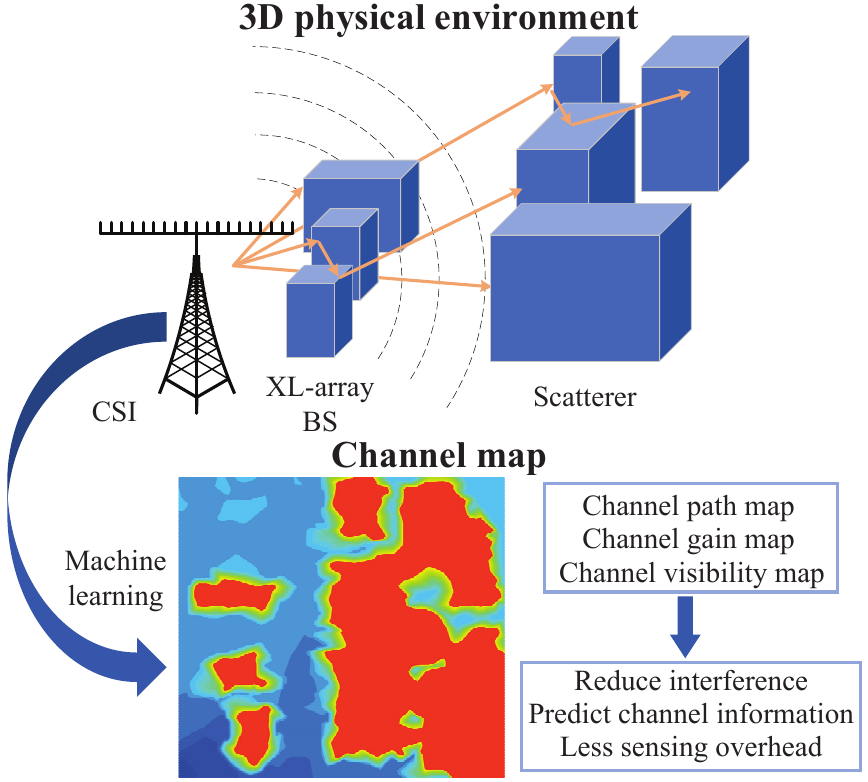}%
		\label{fusion_2}}	
	\caption{{Communication-assisted near-field sensing.}}
	\label{fusion}
\end{figure*}

\subsection{Sensing-Assisted Near-field Resource Allocation}
In addition to single-user communication systems, sensing information/methods can also be used for designing efficient resource allocation in more complex systems.
First, consider a multi-user near-field communication system, where the user location information is acquired by sensing algorithms (e.g., MUSIC). Such information can be  used for designing the XL-array multi-user beamforming without explicit CSI. For example, with user location information, one can effectively reduce inter-user interference by exploiting the beam-focusing effect via concentrating the beam energy at the target near-field user location, thus supporting massive access.
%In addition, user location information can be used for designing near-field multi-access schemes. Specifically, near-field users at different angles and/or ranges generally have small channel correlation due to the beam-focusing effect. Thus, they can be simultaneously scheduled if the inter-user channel correlation is  small enough, which can be estimated from user location information especially in LoS channels. 
For multi-cell communication systems, user location information can be exploited for designing \emph{coordinated} beamforming schemes among neighboring BSs.  For example, if a user is located in the near-field of a serving BS but  in the far-field of a nearby BS in an adjacent cell, analog beamforming of these two BSs can be jointly designed based on the user location information to maximize the received SNR at the user, while reducing interference to other users. However, in practice, there may not always exist LoS channels between the users and XL-array BSs. Therefore, there is a need to design robust beamforming based on user location information to improve reliability.

\section{Communication-Assisted Near-field Sensing}

Besides sensing-assisted near-field communication, communication functionalities in cellular networks can also be used to assist near-field sensing to improve the sensing range, resolution, accuracy, and reliability.

\subsection{Networked Sensing with Information Fusion}
Near-field radar sensing typically relies on the LoS channels between the XL-array and targets, which, however, may not always exist due to random obstacles and environmental scatterers. 
To tackle this issue, a promising approach is leveraging wireless network architectures to achieve near-field \emph{networked sensing}  through information fusion \cite{LiufanISACand6G1}.

First, consider a challenging scenario where the XL-array BS needs to sense a target located beyond its LoS due to obstacles in between, as shown in Fig.~\ref{fusion_1}. 
{An effective solution is delegating the sensing/localization task to a \emph{proxy} XL-array BS in the existing wireless system, which is capable of establishing an LoS link to the target in its near-field, hence improving the resource utilization efficiency as well.} By exploiting spherical wavefront propagation, the proxy BS can efficiently estimate the target range and angle without the need of additional anchors as in the far-field case.  Then, the proxy BS sends its estimated information to the serving BS, which then can estimate the target location based on the relative positions.
%Despite the above advantages, there are still several issues that need to be addressed. For instance, it is important to select a suitable proxy BS  for target sensing. 
Generally speaking, it is desirable to select a proxy BS that is close to the target for more accurate localization, as well as close to the serving BS to achieve more reliable target location estimation.

Next, to further improve the radar sensing performance, an effective approach is to use widely deployed BSs to achieve \emph{near-/mixed}-field networked sensing, so that the BSs can share their obtained angle/range/Doppler information with each other for jointly estimating the locations and velocities of targets, as illustrated in Fig.~\ref{fusion_1}. This approach has several advantages. First, for near-field networked sensing, the target is located within the near-field of multiple BSs. As such, each of these BSs can  sense the target  independently and then send the obtained information to the serving BS to achieve more accurate target localization. Among others, it is necessary to study how to effectively fuse all information to achieve accurate and high-resolution localization by taking into account various factors such as sensing SNR, number of XL-array antennas, spatial correlation of proxy BSs. 
In addition, multiple proxy BSs also enable \emph{multi-view} sensing, which is particularly appealing to extract useful three-dimensional (3D) information of extended targets. Apart from the ground BSs at fixed locations, mobile communication infrastructures, such as unmanned aerial vehicles (UAVs), can also be employed to enable aerial near-field sensing in 3D space \cite{FengzhiyongUAVISAC}. For example, by judiciously designing the UAV trajectory, the target location can be estimated based on multiple reflected signals over its trajectory. In this case,  the UAV trajectory can be optimized to minimize the estimation MSE of near-field sensing, while the coordinate the UAVs and ground BSs can be designed to maximize the sensing range.

\vspace{-5pt}
\subsection{Channel-knowledge-map Assisted Near-field Sensing}

A major challenge in near-field sensing arises from sensing interference caused by environment clutters, which generate non-LoS (NLoS) paths between the XL-array BS and targets. These NLoS paths usually deteriorate radar sensing performance as they provide little useful information of target angle/range/Doppler. {To address this issue, a promising approach is to utilize \emph{channel knowledge maps} (CKMs) to effectively eliminate/cancel the interference from static clutters/scatterers, which can be efficiently updated in real-time by using e.g., predictive update and  incremental learning methods \cite{ZengyongRadiomapping}.} Specifically, CKMs are location-specific datasets (rather than conventional coarse site-specific) that contain useful information regarding intrinsic wireless channels as illustrated in Fig.~\ref{fusion_2}. For near-field communication/sensing, typical examples may include the channel path map, channel gain map, channel \emph{visibility} map, etc. For example, the channel path map aims to predict all essential information of channel paths at each potential user location in the near- or far-field. This is particularly useful for near-field radar sensing as it provides  background noise/clutter information. As such, the XL-array can cancel the resulted interference and then accurately estimate the target angle/range. Second, the channel gain map can be used to reconstruct the layout of  physical environment and environment obstacles. Such information can be employed  to resolve the data association issue in near-field sensing. {Moreover, the channel visibility map characterizes the spatial \emph{non-stationarity} of near-field channels with spatial decorrelation across the array aperture, whereby different users/targets may observe different regions of the XL-array.} From a communication perspective, the channel visibility map is useful for designing efficient user scheduling schemes to reduce the inter-user interference by using, for example, the array-division method. While for radar sensing, the channel visibility map implicitly provides useful information of environment scatterers/obstacles. With such information, the XL-array does not need to scan the entire angular and range domains for localization. Instead, a more efficient approach is that each sub-array independently scans the targets within its visibility region only, thereby greatly reducing the sensing subspace and hence the sensing overhead. In addition, the sensing information from different sub-arrays can be fused together to further improve  radar sensing accuracy and resolution.
\vspace{-5pt}

% \usepackage{multirow}%no need multirow！
\begin{table*}[]\centering
\belowrulesep=0pt
\aboverulesep=0pt

	\caption{{Summary  of new design issues for near-field ISAC.}}
	{\begin{tabular}{c|c|c}
		\toprule[1.5pt]
		Main sections	& Different approaches &  New design issues \\\midrule[1.2pt]
		\multirow{3}{*}{\makecell[c]{Near-field joint \\communication and sensing}} &\makecell[c]{Near-field JC\&S in \\ narrow-band systems} & \multicolumn{1}{l}{\begin{tabular}[c]{@{}l@{}}\textbullet~Near-field multi-beam design for JC\&S\\ \textbullet~Near-field target tracking\end{tabular}} \\ \cmidrule{2-3}
		& \makecell[c]{Near-field JC\&S in\\ wide-band systems}& \multicolumn{1}{l}{\begin{tabular}[c]{@{}l@{}}\textbullet~TDD based near-field JC\&S \end{tabular}}\\ \midrule[1pt]
		\multirow{5}{*}{\makecell[c]{Sensing assisted\\near-field communication}} & \makecell[c]{Sensing-assisted near-field\\beam training/tracking} & \multicolumn{1}{l}{\begin{tabular}[c]{@{}l@{}@{}@{}@{}}\textbullet~Anchor assisted near-field beam training\\\textbullet~Radar sensing assisted near-field beam tracking\end{tabular}} \\ \cmidrule{2-3} 
		& \makecell[c]{Attitude information-assisted\\near-field beam focusing} & \multicolumn{1}{l}{\begin{tabular}[c]{@{}l@{}@{}}\textbullet~Beam-focusing design based on  attitude information\\\textbullet~Local-global coordinate near-field ISAC system\\\end{tabular}} \\ \cmidrule{2-3} 
		& \makecell[c]{Sensing-assisted near-field\\resource allocation} & \multicolumn{1}{l}{\begin{tabular}[c]{@{}l@{}@{}}\textbullet~Near-field multi-user interference elimination\\\textbullet~Near-/Mixed-field coordinated beamforming\end{tabular}}\\ \midrule[1pt]
		\multirow{3}{*}{\makecell[c]{Communication assisted \\ near-field sensing}} & \makecell[c]{Networked sensing with\\information fusion} &\multicolumn{1}{l}{\begin{tabular}[c]{@{}l@{}@{}}\textbullet~Beyond-LoS near-field sensing\\\textbullet~Multi-BS assisted networked near-field sensing\end{tabular}} \\ \cmidrule{2-3} 
		& \makecell[c]{CKM assisted\\near-field sensing} &  \multicolumn{1}{l}{\begin{tabular}[c]{@{}l@{}@{}}\textbullet~Channel path map assisted radar sensing\\\textbullet~Channel visibility map assisted localization\end{tabular}} \\ \midrule[1pt]
		\multirow{4}{*}{\makecell[c]{Future directions}} & \makecell[c]{IRS assisted near-field ISAC} & \multicolumn{1}{l}{\begin{tabular}[c]{@{}l@{}@{}@{}@{}}\textbullet~XL-IRS placement for near-field 
 C\&S\end{tabular}} \\ \cmidrule{2-3} 
		& \makecell[c]{DL for near-field ISAC} & \multicolumn{1}{l}{\begin{tabular}[c]{@{}l@{}@{}}\textbullet~DL enabled hybrid beamforming\\\textbullet~Few-shot learning based near-field sensing\end{tabular}} \\ \cmidrule{2-3} 
		& \makecell[c]{MA/FA-enabled near-field ISAC} & \multicolumn{1}{l}{\begin{tabular}[c]{@{}l@{}@{}}\textbullet~Antenna position optimization for near-field ISAC \end{tabular}}\\ \bottomrule[1.5pt]	
	\end{tabular}}
 \label{TaIssue}
\end{table*}

\section{Conclusions and Future Directions}

In this article, we provided an overview of near-field ISAC for XL-array wireless systems. Specifically, we first introduced the unique near-field channel characteristics and appealing advantages of near-field C\&S. Then, we presented three approaches for near-field ISAC and discussed their new design challenges and promising solutions as summarized in Table~\ref{TaIssue}. Three other directions for future research in near-field ISAC are outlined below.

\vspace{-7pt}
\subsection{Intelligent Reflecting Surface-Assisted Near-field ISAC}
Extremely large-scale intelligent reflecting surface (XL-IRS) has emerged as a promising technology to dynamically reconfigure wireless propagation environment \cite{WCMadd1}, for which communication users and/or sensing targets are very likely to be located in the near-field region of an IRS. 
Among others, it is important to study how to deploy XL-IRSs in near-field wireless systems. 
For wireless communication, the BS-side XL-IRS placement can provide a higher multiplexing gain than the far-field case due to the higher channel rank, while the user-side XL-IRS placement may help mitigate inter-user interference by exploiting the beam-focusing effect. On the other hand, for radar/wireless sensing, XL-IRS placement at the BS-/target-side which usually operates on the reflect mode, may affect the sensing coverage and accuracy. 
%Besides, to balance the C\&S performance, the active beamforming of the XL-array BS and passive beamforming of the XL-IRS need to be jointly designed, which becomes more complicated in wide-band systems due to the near-field beam-split effect. 

\vspace{-5pt}
 \subsection{Deep Learning for Near-field ISAC}
 Deep learning (DL) methods can be used to address several issues in near-field ISAC, as discussed below. First, for the communication design, customized deep neural networks can be designed to reduce the computational complexity of the hybrid/digital beamforming and channel estimation designs for XL-arrays.
On the other hand, for near-field sensing, DL methods such as few-shot learning can be invoked for designing near-field predictive beam tracking with a small number of echo signals only, hence greatly improving the beam tracking efficiency and reliability, while reducing the design complexity. 
 
\subsection{Movable/Fluid Antenna Enabled Near-field ISAC}
To reduce the hardware cost of XL-arrays, movable/fluid antennas (MAs/FAs) can be employed to enable near-field ISAC with a small number of antennas only, by effectively forming virtual large-aperture arrays. For the MA-enabled near-field communications, the positions of antennas  can be jointly optimized to achieve the near-field beam-focusing gain, while eliminating interferences of the near-field grating-lobes arising from sparsely positioned antennas. In addition, the flexible inter-antenna spacing of MAs can be exploited to design efficient near-field multi-beam training with low overhead. 
For near-field sensing, one can form different structures of arrays (e.g., coprime and nested arrays) by optimizing the antenna positions. Hence, it is necessary to design efficient near-field angle-and-range estimation algorithms for different forms of arrays as well as study their sensing performance. % Moreover, DL methods can also be applied for near-field gesture recognition and motion prediction applications, which essentially exploits the near-field received signal features for estimation/recognition.

\bibliographystyle{IEEEtran}
\bibliography{ISAC_magazine.bib}

\end{document}